\documentclass[12pt]{article}
\begin{document}
\title{Planck Oscillators in the Background Dark Energy$*$}
\author{B.G. Sidharth\\
International Institute for Applicable Mathematics \& Information Sciences\\
Hyderabad (India) \& Udine (Italy)\\
B.M. Birla Science Centre, Adarsh Nagar, Hyderabad - 500 063
(India)}
\date{}
\maketitle \footnote{$^*$Based on the Paper at the Max Born
Symposium, 2009, Wroclaw}
\begin{abstract}
We consider a model for an underpinning of the universe: there are
oscillators at the Planck scale in the background dark energy.
Starting from a coherent array of such oscillators it is possible to
get a description from elementary particles to Black Holes including
the usual Hawking-Beckenstein theory. There is also a description of
Gravitation in the above model which points to a unified description
with electromagnetism.
\end{abstract}
\section{Introduction}
Max Planck, more than a century ago introduced a combination of the
well known fundamental constants, $\hbar,G,c$ that gave a length,
mass and time scale viz.,
$$l = \sqrt{\frac{\hbar G}{c^3}} \sim 10^{-33}cm$$
\begin{equation}
m = \sqrt{\frac{\hbar c}{G}} \sim 10^{-5}gm\label{e1}
\end{equation}
$$t = \sqrt{\frac{\hbar G}{c^5}} \sim 10^{-42}sec$$
We can easily verify that $l$ plays the role of the Compton length
and the Schwarzchild radius of a black hole of the mass $m$
\cite{kiefer}
\begin{equation}
l = \frac{\hbar}{2mc}, \, l = \frac{2Gm}{c^2}\label{e2}
\end{equation}
Today in various Quantum Gravity approaches including String theory,
the Planck length $l$ is considered to be the fundamental minimum
length, and so also the time interval $t$.
\section{Quantum Strings}
In spite of great success, the standard theory has failed to
quantize gravitation. One of the obstacles has been the point
spacetime concept ingrained in these theories leading to infinities.
For the past few decades Quantum Gravity schemes as also String
theory have attempted to break out of this limitation. Let us first consider string theory.\\
We begin with the important work of T. Regge in the fifties
\cite{reg,roma,tassie}, in which he mathematically analysed using
techniques like analytically continuing the angular momentum into
the complex plane, particle resonances. These resonances seem to
fall along a straight line plot, with the angular momentum being
proportional to the square of the mass.
\begin{equation}
J \propto M^2,\label{ex}
\end{equation}
All this suggested that resonances had angular momentum, on the one hand and resembled extended objects, that is particles smeared out in space.
For example, mathematically this was like two heavy objects attached to the two ends of a string, or a rotating stick.\\
This went contrary to the belief that truly elementary particles
were points in space. In fact at the turn of the twentieth century,
Poincare, Lorentz, Abraham and others had toyed with the idea that
the electron had a finite extension, but they had to abandon this
approach, because of a conflict with Special Relativity. The problem
is that if there is a finite extension for the electron then
forces on different parts of the electron would exhibit a time lag, requiring the so called Poincare stresses for stability \cite{rohr,barut,feynman}.\\
In this context, it may be mentioned that in the early 1960s, Dirac came up with an imaginative picture of the electron, not so
much as a point particle, but rather a tiny closed membrane or bubble. Further, the higher energy level oscillations of this
membrane would represent the ``heavier electrons'' like muons \cite{dirac}.\\
Then, in 1968, G. Veneziano came up with a unified description of the Regge resonances (\ref{ex}) and other scattering processes.
Veneziano considered the collision and scattering process as a black box and pointed out that there were in essence, two
scattering channels, $s$ and $t$ channels. These, he argued gave a dual description of the same process \cite{ven,venezia}.\\
In an $s$ channel, particles A and B collide, form a resonance which quickly disintegrates into particles C and D. On the
other hand we have in a $t$ channel scattering  particles A and B approach each other, and interact via the exchange of a
particle $q$. The result of the interaction is that particles C and D emerge. If we now enclose the resonance and the
exchange particle $q$ in an imaginary black box, it will be seen that the $s$ and $t$ channels describe the same input
and the same output: They are essentially the same.\\
There is another interesting hint which we get from Quantum Chromo
Dynamics. Let us come to the inter-quark potential
\cite{lee,smolin}. There are two interesting features of this
potential. The first is that of confinement, which is given by a
potential term like
$$V (r) \approx \sigma r, \quad r \to \infty ,$$
where $\sigma$ is a constant. This describes the large distance behavior between two quarks. The confining potential
ensures that quarks do not break out of their bound state, which means that effectively free quarks cannot be observed.\\
The second interesting feature is asymptotic freedom. This is
realized by a Coulumbic potential
$$V_c (r) \approx - \frac{\propto (r)}{r} (\mbox{small}\, r)$$
$$\mbox{where}\, \propto (r) \sim \frac{1}{ln(1/\lambda^2r^2)}$$
The constant $\sigma$ is called the string tension, because there are string models which yield $V(r)$. This is because, at
large distances the inter-quark field is string like with the energy content per unit length becoming constant. Use of the
angular momentum - mass relation indicates that $\sigma \sim (400 MeV)^2$.\\
Such considerations lead to strings which are governed by the
equation \cite{walter,fog,bgskluwer,st}
\begin{equation}
\rho \ddot {y} - T y'' = 0,\label{ae1}
\end{equation}
\begin{equation}
\omega = \frac{\pi}{2l} \sqrt{\frac{T}{\rho}},\label{ae2}
\end{equation}
\begin{equation}
T = \frac{mc^2}{l}; \quad \rho = \frac{m}{l},\label{ae3}
\end{equation}
\begin{equation}
\sqrt{T/\rho} = c,\label{ae4}
\end{equation}
$T$ being the tension of the string, $l$ its length and $\rho$ the
line density and $\omega$ in (\ref{ae2}) the frequency. The
identification (\ref{ae2}),(\ref{ae3}) gives (\ref{ae4}), where $c$
is the velocity of light, and (\ref{e1}) then goes over to the usual
d'Alembertian or massless Klein-Gordon equation. (It is worth noting
that as $l \to 0$ the potential energy
which is $\sim \int^l_0 T (\partial y/\partial x)^2 dx$ rapidly oscillates.)\\
Further, if the above string is quantized canonically, we get
\begin{equation}
\langle \Delta x^2 \rangle \sim l^2.\label{ae5}
\end{equation}
The string effectively shows up as an infinite collection of
harmonic oscillators \cite{fog}. It must be mentioned that
(\ref{ae5}) and (\ref{ae2}) to (\ref{ae4}) both show that $l$ is of
the order of the Compton wavelength. This has been called one of the
miracles of string theory by Veneziano \cite{veneziano}. In fact the
minimum length $l$ turns out to be given by $T/\hbar^2 = c/l^2$,
which from (\ref{ae3}) and (\ref{ae4}) is seen to give the Compton wavelength.\\
This is a description of what may be called a ``Bosonic String''. These theories have certain technical problems, for example they
allow the existence of tachyons. Further they do not easily meet the requirements of Quantum theory, as for example the commutation
relations. The difficulties are resolved only in twenty six dimensions.\\
If the relativistic quantized string is given rotation
\cite{ramond}, then we get back the equation for the Regge
trajectories given in (\ref{ex}) above. Here we are dealing with
objects of finite extension rotating with the velocity of light
rather like spinning black holes. It must be pointed out that, in
superstring theory, there is an additional term $a_0$
\begin{equation}
J \leq (2\pi T)^{-1} M^2 + a_0 \hbar, \, \mbox{with}\, a_0 = +1 (+2)
\, \mbox{for \, the \, open\, (closed)\, string.}\label{ae6}
\end{equation}
In equation (\ref{ae6}) $a_0$ comes from a zero-point energy effect.
When $a_0 = 1$ we have the usual gauge bosons and when $a_0 = 2$
we have the gravitons.\\
The theory of Quantum Super Strings in contrast requires only ten dimensions. Here, Quantum operators describing anti-commuting
variables satisfy anti-commutation relations. Indeed this bivalence is a hallmark of supersymmetry itself.\\
The extra dimensions that appear in String theories reduce to the four dimensions of the physical spacetime by virtue of the fact
that the redundant dimensions are treated as curled up into a negligible extension, in the manner suggested by Kaluza and later
Klein in the early twentieth century. Kaluza's original motivation had been to unify electromagnetism and gravitation by introducing
a fifth negligible coordinate. The curling up takes place at the Planck scale \cite{kaluza}.\\
A finite extension for an elementary particle, as in String theories
can be shown to lead to new commutation relations, as was done by
Snyder in the forties. In this case two space coordinates like $x$
and $y$ do not commute. Snyder's original motivation had been to
fudge and eliminate singularities and divergences in Quantum fields.
We remark that what this implies is that space coordinates in some
sense take on the mathematical character of momenta in addition,
though this happens at very small scales or high energies.
Effectively there is a modification of the Uncertainty Principle
\begin{equation}
\Delta x \geq \frac{\hbar}{\Delta_P} + l^2
\frac{\Delta_P}{\hbar}\label{ae7}
\end{equation}
What all this means is we cannot go down to lower and lower space
scales arbitrarily. As we approach the minimum length we return to
the larger
universe \cite{witten}.\\
The interesting thing about Quantum Superstring theory is the
natural emergence of the spin 2 graviton as can be seen from
(\ref{ae6}), or
as Witten puts it, the theory ``predicts'' gravitation.\\
Meanwhile Supersymmetry or SUSY developed in parallel. This theory requires that each particle with integral spin has a counterpart with
the same mass but having half integral spin. That is Bosons have their supersymmetric counterparts in Fermions. SUSY is then broken
so that the counterparts would have a much greater mass, which would then account for the fact that these latter have not been
observed. Nevertheless the fact that in this theory gravitation can be unified with the other forces makes it attractive.\\
Infact this had lead to Supergravity in which the spin 2 graviton has the spin 3/2 counterpart, the gravitino. Supergravity
requires eleven spacetime dimensions, one more than Superstring theory.\\
Unfortunately Supergravity began to fade from the mid eighties because of the fact that, as shown by Witten and others,
handedness cannot easily emerge on reduction to the four physical spacetime dimensions from eleven. On the other hand the
Quantum Super String theory was in comparison altogether more satisfactory. We could say that the earlier Bosonic String
theory worked in a spacetime that was Bosonic, there being no place for spin. QSS works in a Fermionic spacetime where
we have the modification (\ref{ae7}).\\
So in the mid eighties ten dimensional QSS displaced Supergravity. There were five QSS theories - $E_8 \times E_8$ heterotic,
$SO(32)$ heterotic, the Type I, the Type IIA and Type IIB. Of these the Type I is an open string while the others form closed
loops. The $E_8 \times E_8$ appeared to explain many features of elementary particles and their forces.\\
However there were some disturbing questions. Why were there five
different theories? After all we need a unique theory. And then why
ten dimensions, while Supersymmetry allows eleven dimensions?
Another not very convincing factor was the fact that particles were
being represented as one dimensional strings. Surely a more general
formulation would have two dimensional surfaces or membranes or even
p-dimensional entities which we may call p-branes. This
generalization resembles the earlier attempt of Dirac's representing
particles as a shell or membrane. Infact if the radius of the circle
shrinks, the mebrane begins to resemble a rolled up object in ten dimensions. It reduces to a Type IIA Superstring.\\
In such deformations certain topological properties can remain conserved. A good example is a knot in a set of field lines.
Such knots or solitons remain as such and exhibit a particle type behaviour. A magnetic monopole can be characterized in
this way, that is as a twisted knot of magnetic lines. It can be said to carry a topological charge. This is to be contrasted
with the charges carried by particles like electrons and quarks which can be put within the framework of the Noether Conservation
Theorem. In this context an interesting conjecture is that of Montonen and Olive \cite{olive}: There could be a dual formulation
in which the roles of the usual charges and topological charges are reversed. In such a formulation for example a particle with
charge $e$ would show up as a soliton with charge $\frac{1}{e}$.\\
Over the past few years, a variant called $M$ Theory arising from these generalizations has attracted much attention. This
theory also uses Supersymmetry, which is broken so that the postulated particles do not have the same mass as the known particles.
Further these new masses must be much too heavy to be detected by current accelerators. The advantage of Supersymmetry is that a
framework is now available for the unification of all the interactions including gravitation. It may be mentioned that under a
SUSY transformation, the laws of physics are the same for all observers, which is the case in General Relativity (Gravitation)
also. Under SUSY there can be a maximum of eleven dimensions, the extra dimensions being curled up as in Kaluza-Klein theories.
In this case there can only be an integral number of waves around the circle, giving rise to particles with quantized energy.
However for observers in the other four dimensions, it would be quantized charges, not energies. The unit of charge would depend
on the radius of the circle, the Planck radius yielding the value $e$. This is the root of the unification of electromagnetism
and gravitation in these theories.\\
The relevance of all this is that p-branes can be characterized as solitons. For example a ten dimensional string can show up
as a p-brane with $p = 5$. In this case a strongly interacting string would be the dual of a weakly interacting 5-brane. In 1990
the Montonen-Olive duality which was between electricity and magnetism in ordinary four dimensional space, was generalized to
four dimensional Superstrings.\\
This duality was called S-duality, to distinguish it from the well
known T-duality which relates two kinds of particles that arise when
the string loops around by a compact dimension: There would be
vibrations on the one hand and multiple windings on the other.
Winding particles over a circle of radius $r$ correspond to
vibrating particles in a circle of radius $1/r$ and conversely on
the lines of (\ref{ae7}). Such a behaviour is characteristic of
minimum spacetime intervals. In this picture the solitonic
interaction
is given by the reciprocal of the string interaction, in confirmity with the Montonen-Olive conjecture.\\
A further interesting development was the realization that in the reduction of the dimensions of spacetime to four dimensions the
string and the corresponding soliton each acquire a T-duality. Moreover the T-duality of the solitonic string is the S-duality of
the fundamental string and conversely. We have here a duality of dualities. It also implies that the interaction charges in one
universe show up as sizes in the dual.\\
Further the eleventh and extra dimension of the M-Theory could be
shrunk, so that there would be two ten dimensional universes
connected by the eleven dimensional spacetime. Now particles and
strings would exist in the parallel universes which can interact
through gravitation. The interesting aspect of the above scenario is
that it is possible to concieve of all the four interactions
converging at an energy far less than the Planck energy
$(10^{19}GeV$). Infact the Planck energy is so high that it is
beyond forseeable experiments. Thus this would bring the eleven
dimensional M-Theory closer to experiment. There have been further
developments involving what are called Dirichlet surfaces. It is now
suspected that black holes can be treated as intersecting black
branes wrapped around seven curled up dimensions. There is here, an
interesting interface between M-Theory and black hole physics
\cite{green}. In M-Theory, the position coordinates become matrices
and this leads to, a noncommutative geometry or fuzzy spacetime in
which spacetime points are no longer well defined \cite{madore}
$$[x,y] \ne 0$$
From this point of view the mysterious $M$ in M-Theory could stand
for Matrix, rather than Membrane.\\
So M-Theory is the new avatar of QSS. Nevertheless it is still far
from being the last word. There are still any number of routes for
compressing ten dimensions to our four dimensions. There is still no
contact with experiment. It also appears that these theories lead to
an unacceptably high cosmological constant and so on.\\
To bypass these difficulties, string theorists have had to invoke
the concept of a landscape of universes together with the anthropic
principle. The idea is that each of the $100^{500}$ or so solutions
represents a universe, each with its own characteristic values for
physical constants. The anthropic principle is then invoked to
explain why our universe has the observed values for the physical
constant, and this includes the cosmological constant. All this
however has not gone well with many physicists and the entire
spectrum of string theory has come under severe criticism in the
past few years \cite{laughlin,smolin2,bgsuturn}. Even prominent
string theorists like David Gross now express pessimism about string
theory. As Susskind puts it, \cite{susskind} "Confusion and
disorientation reign; cause and effect break down; certainty
evaporates; all the old rules fail. That's what happens when the
dominant paradigm breaks down." Let us explore further, given the
above context.
\section{The Planck Oscillators}
Spacetime intervals smaller than given in (\ref{e1}) and (\ref{e2})
are meaningless both classically and Quantum mechanically.
Classically because we cannot penetrate the Schwarzchild radius, and
Quantum mechanically because we encounter unphysical phenomena
inside a typical Compton scale. All this has been discussed in
greater detail in the literature (Cf.ref.\cite{tduniv} and several
references therein). We could of course go to smaller intervals by
abandoning the Planck mass and the fundamental constants in
(\ref{e1}) and (\ref{e2}) -- we will come back to this point a
little later. In any case, it is worth pointing out that Quantum
mechanically it is meaningless to speak about spacetime points, as
these would imply infinite momenta and energy. This is at the root
of the infinities and divergences that we encounter, both in the
classical theory of the electron as also in Quantum mechanics and
Quantum Field Theory. In Quantum Field Theory we have to take
recourse to the mathematical device of Renormalization
to overcome this difficulty.\\
At another level, it may be mentioned that the author's 1997 model
invoked a background dark energy and fluctuations therein to deduce
a model of the universe that was accelerating with a small
cosmological constant, together with several other relations
completely consistent with Astrophysics and Cosmology
(Cf.ref.\cite{bgscu} and several references therein). At that time
it may be recalled, the accepted standard big bang model told us
that the universe was dominated by dark matter and was consequently
decelerating and would eventually come to a halt. However the
observations of distant supernovae by Perlmutter and others
confirmed in 1998 the dark energy driven accelerating universe of
the author. All this is well known.\\
It is against this backdrop and the difficulties with Quantum
Gravity approaches as detailed in Section 2, that the author had put
forward his model of Planck oscillators in the dark energy driven
Quantum vacuum, several years ago (Cf.ref.\cite{uof} and several
references therein, \cite{fpl2000} and \cite{ijtp}). To illustrate
this model let us consider an array of $N$ particles, spaced a
distance $\Delta x$ apart, which behave like oscillators that are
connected by springs. As is known we then have
\cite{uof,bgsfpl15,good,vandam} (Cf.in particular ref.\cite{vandam})
$$r = \sqrt{N \Delta x^2}$$
\begin{equation}
ka^2 \equiv k \Delta x^2 = \frac{1}{2} k_B T\label{e3}
\end{equation}
where $k_B$ is the Boltzmann constant, $T$ the temperature, $r$ the
total extension and $k$ is the spring constant given by
\begin{equation}
\omega_0^2 = \frac{k}{m}\label{e4}
\end{equation}
\begin{equation}
\omega = \left(\frac{k}{m} a^2\right)^{\frac{1}{2}} \frac{1}{2} =
\omega_0 \frac{a}{r}\label{e5}
\end{equation}
It must be pointed out that equations (\ref{e3}) to (\ref{e5}) are
quite general and a part of the well known theory referred to in
\cite{bgsfpl15,good,vandam}. In particular there is no restriction
on the temperature $T$. $m$ and $\omega$ are the mass of the
particle and frequency of oscillation. In (\ref{e4}) $\omega_0$ is
the frequency of the individual oscillator, while in (\ref{e5})
$\omega$ is the frequency of the array of $N$ oscillators, $N$
given in (\ref{e3}).\\
We now take the mass of the particles to be the Planck mass and set
$\Delta x \equiv a = l$, the Planck length as the mass and length
are free parameters. In other words, instead of considering a single
Planck oscillator as in String theory, we are now considering a
coherent array of such oscillators, rather like coherent vibrating
atoms in a linear crystal. We also use the well known Einstein-de
Broglie relations that give quite generally the frequency in terms
of energy and mass.
\begin{equation}
E = \hbar \omega = mc^2\label{e6}
\end{equation}
It may be immediately observed that if we use (\ref{e4}) and
(\ref{e3}) we can deduce that
$$k_B T \sim mc^2$$
Independently of the above steps this agrees with the (Beckenstein)
temperature of a Black Hole of Planck mass in the usual theory.
Indeed as noted, Rosen \cite{rosen} had shown that a Planck mass
particle at the Planck scale can be considered to be a Universe in
itself with a
Schwarzchild radius equalling the Planck length.\\
Thus we have shown from the above theory of oscillators that an
oscillator with the Planck mass and with a spatial extent at the
Planck scale has the same temperature as the Beckenstein temperature
of a Schwarzchild Black Hole of mass given by the Planck mass. The
above results can also be obtained by a different route as described
in \cite{bgsijmpa}.
\section{Elementary Particles and Black Hole Thermodynamics}
We have also argued elsewhere that, given the well known effect that
the universe consists of $N \sim 10^{80}$ elementary particles like
the pion, it is possible to deduce that a typical elementary
particle consists of $n \sim 10^{40}$ Planck oscillators. As this
has been discussed extensively in the references given, we merely
quote the final result. These form a coherent array of $n$ elements
described by equations (\ref{e3}) to (\ref{e6}) above. In this case
$N$ in (\ref{e3}) becomes $n$ and we can immediately deduce the
following
\begin{equation}
l_\pi = \sqrt{n}l \, \, \, m_\pi = \frac{m}{\sqrt{n}}\label{A}
\end{equation}
which give the Compton wavelength and mass of a typical elementary
particle represented by $l_\pi$ and $m_\pi$. So a typical elementary
particle is given as the lowest energy state of the above coherent
array of $n$ Planck
oscillators.\\
Interestingly the above description can lead to an immediate
correspondence with black hole thermodynamics. We now rewrite
equation (\ref{e5}) as, (interchanging the roles of $\omega$ and
$\omega_0$),
$$\omega_0 = \frac{r}{a} \omega$$
Remembering that, quite generally, the frequency and mass are
related by (\ref{e6}), i.e.,
$$\omega = \frac{mc^2}{\hbar},$$
we get on using (\ref{e3})
\begin{equation}
\hbar \omega \langle \frac{l}{r}\rangle^{-1} \approx mc^2 \times
\frac{r}{l} \approx Mc^2 = \sqrt{\bar{N}} mc^2\label{e7}
\end{equation}
where we now consider not the lowest energy states of the array as
previously but rather energy states much higher than the Planck
energy. Generally, if an arbitrary mass $M$, as in (\ref{e7}), is
given in terms of $\bar{N}$ Planck oscillators, in the above model,
then we have from (\ref{e7}) and (\ref{e3}):
\begin{equation}
M = \sqrt{\bar{N}} m \, \mbox{and} \, R = \sqrt{\bar{N}}
l,\label{e8}
\end{equation}
where $R$ is the radius or extension of the object. We must stress
the factor $\sqrt{\bar{N}}$ in (\ref{e8}), arising from the fact
that the oscillators are coupled, as given in (\ref{e3}). If the
oscillators had not been coupled, or equivalently had not formed a
coherent system, then we would have, for example, $M = \bar{N}m$ or
$R = \bar{N}l$ instead of (\ref{e8}). Using the fact that $l$ has
been chosen to be the Schwarzchild radius of the (Planck) mass $m$,
this gives immediately,
$$R = 2GM/c^2$$
This shows that if an arbitrary mass $M$ consists of $\bar{N}$
coherent Planck oscillators as above, and specifically equation
(\ref{e8}), then its radius $R$ is given by the above expression,
which is its Schwarzchild radius. In other words, such an object
shows up as a Schwarzchild Black Hole. It must be emphasized that
the expression for $R$ follows from the theory of oscillators,
specifically equation (\ref{e8}) and shows that it is identical to
the Schwarzchild radius for the same mass $M$. We have merely used
the known equivalence of the Planck length and Schwarzchild radius
for the Planck mass.
\section{Thermodynamic Gravitation}
We can push the above consideration further. So far we have
considered only a coherent array \cite{ijtp2}. This is necessary for
meaningful physics and leads to the elementary particle masses and
their other parameters as seen above. Cercignani \cite{cer} had used
Quantum oscillations, though just before the dark energy era --
these were the usual Zero Point oscillations, which had also been
invoked by the author in his model. Invoking gravitation, what he
proved was, in his own words, ''Because of the equivalence of mass
and energy, we can estimate that this (i.e. chaotic oscillations)
will occur when the former will be of the order of $G[(\hbar \omega
)c^{-2}]^2 [\omega^{-1}c]^{-1} = G\hbar^2\omega^3 c^{-5}$, where $G$
is the constant of gravitational attraction and we have used as
distance the wavelength. This must be less than the typical
electromagnetic energy $\hbar \omega$. Hence $\omega$ must be less
than $(G\hbar)^{-1/2}c^{5/2}$, which gives a gravitational cut off
for
the frequency in the zero-point energy."\\
In other words he deduced that there has to be a maximum cut off
frequency of oscillators given by
\begin{equation}
G\hbar \omega^2_{max} = c^5\label{e10}
\end{equation}
for the very existence of coherent oscillations. We would like to
point out that if we use the well known equation encountered above
namely
$$\hbar \omega = mc^2,$$
in equation (\ref{e10}) we get the well known relation
\begin{equation}
Gm^2_P \approx \hbar c\label{B}
\end{equation}
which shows that at the Planck scale the gravitational and
electromagnetic strengths are of the same order. This is not
surprising because it was the very basis of Cercignani's derivation
-- if indeed the gravitational energy is greater than that given in
(\ref{B}) that is greater than the electromagnetic energy, then the
Zero Point oscillators, which we have called the Planck oscillators
would become chaotic and incoherent -- there would be no physics.\\
Let us now speak only in terms of the background dark energy. Then
we can argue that (\ref{B}) is the necessary and sufficient
condition for coherent Planck oscillators to exist, in order that
there be elementary particles as given by (\ref{A}) and the rest of
the requirements for the meaningful physical universe. In other
words gravitational energy represented by the gravitation constant
$G$ given in (\ref{B}) is a measure of the energy from the Quantum
background that allows a physically meaningful universe -- in this
sense it is
not a separate fundamental interaction. We will return to this point.\\
It is interesting that (\ref{B}) also arises in Sakharov's treatment
of gravitation where it is a residual type of a zero point energy
\cite{sakharov,tduniv}.\\
To proceed if we use (\ref{A}) in (\ref{B}) we can easily deduce
\begin{equation}
Gm^2 \approx \frac{e^2}{n} = \frac{e^2}{\sqrt{N}}\label{C}
\end{equation}
where now $N \sim 10^{80}$, the number of particles in the
universe.\\
Equation (\ref{C}) has been known for a long time emperically, as an
accident without any fundamental explanation. Here we have deduced
it on the basis of the Planck oscillator model. Equation (\ref{C})
too brings out the relation between gravitation and the background
Zero Point Field or Quantum vacuum or dark energy. It shows that the
gravitational energy has the same origin as the electromagnetic
energy but is in a sense a smeared out effect over the $N$ particles
of the universe.  We will argue in Section 7 that the smearing out
is due to the fact that we require an array of oscillators. In the
context of the above considerations we can now even claim that
(\ref{C}) gives the desired unified description of electromagnetism
and gravitation and not an ad hoc formula.
\section{Black Holes Again}
If we use (\ref{e4}), (\ref{e3}) and (\ref{e10}) we get \cite{ijtp}
$$k_B T = m \omega^2_{max} l^2 = \frac{c^5/G \hbar}{ml^2} = \frac{\hbar
c^3}{Gm},$$ remembering that $l$ by (\ref{e2}) is also the Compton
wavelength. That is we get
\begin{equation}
k_B T = \frac{\hbar c^3}{Gm}\label{e11}
\end{equation}
Equation (\ref{e11}) is the well known Beckenstein temperature
formula valid for a Black Hole of arbitrary mass but derived here
for the Planck mass.\\
Can we now generalize equation (\ref{e11}) to the case of a Black
Hole of arbitrary mass, as in the original Beckenstein formula but
using only the characterization of the Black Hole in terms of Planck
oscillators, as above? This is what we will do. In fact to a Black
Hole of mass $M$ characterized in terms of $\bar{N}$ oscillators as
in equation (\ref{e8}), we associate a Black Hole temperature
defined by
$$\bar{T} = \frac{T}{\sqrt{\bar{N}}},$$
where $T$ is given in (\ref{e11}). ($\bar{N}$ here is not the number
of particles in the universe). Using this with (\ref{e8}) in
(\ref{e11}) we immediately get
\begin{equation}
k_B \bar{T} = \frac{\hbar c^3}{GM}\label{e12}
\end{equation}
Equation (\ref{e12}) which is the analogue of (\ref{e11}) is the
required result. After this identification, we next use the
following known relations for a Schwarzchild Black Hole
\cite{ruffinizang}:
\begin{equation}
dM = TdS, S = \frac{kc}{4\hbar G}A,\label{e13}
\end{equation}
where $T$ is the Black Hole temperature, now identified with
(\ref{e12}), $S$ the entropy and $A$ is the area of the Black Hole.
The area is given by, using (\ref{e8})
\begin{equation}
A = \bar{N} l^2\label{e14}
\end{equation}
because, this area is $\sim R^2$. Alternatively this shows that
there are $\bar{N}$ elementary areas $l^2$ forming the Black Hole.
Indeed this defines the basic quantum of area of quantum gravity
approaches and is in pleasing agreement with the result of Baez
deduced from a
different quantum gravity consideration \cite{baez}.All this also answers the 'tHooft conjecture about the black hole-elementary particle mass specturm.\\
Using equations (\ref{e8}), (\ref{e11}) and (\ref{e14}), we can
easily see that equation (\ref{e13}) is valid for the mass $M$ given
by (\ref{e8}) or (\ref{e7}).\\
This completes the identification of Black Holes characterized by
coherent Planck oscillators, with the conventional
Hawking-Beckenstein theory.
\section{Remark}
As already noted, in one sense, we can get lengths $< l_P$ if the
mass $< m_P$ though such a scale would no longer be in terms of the
fundamental constants, unlike the Planck scale (\ref{e1}). However,
let us consider the following relation (Cf.ref.\cite{tduniv}),
\begin{equation}
\omega^2_{max} = \frac{c^2}{l^2}\label{e17}
\end{equation}
This follows from the theory of phonons in an array of coherent
oscillators e.g. atoms in a linear crystal  as in our model. If we
use (\ref{e17}) in (\ref{e10}), then we get the Planck length
(\ref{e1}). In other words, the Planck length is the result of not
just a single oscillator but rather a whole array of oscillators as
in our theory. A small scale would lead to an unphysical chaotic universe.\\
It is no longer arbitrarily prescribed as in (\ref{e1}). From this
point of view, there is a distinction in the interpretation of
gravitation as compared to Sakharov's formulation alluded to. True
gravitation shows up as a residual energy according to (\ref{B}) as
in Sakharov's theory. But now, this is due to the result of the
array of oscillators at the Planck scale in the background dark
energy.
\section{Conclusion}
We have shown that it is possible to consider the universe to have
an underpinning of Planck oscillators in the background dark energy.
This leads to a meaningful description of the universe of elementary
particles and also of  black hole thermodynamics. Finally it
provides a description of gravitation, not as a separate fundamental
interaction, but rather as the residual energy of the background
dark energy that is a result of the fact that there is a minimum
fundamental spacetime interval that is required for a meaningful
universe.

\end{document}